\documentclass[12pt,superscriptaddress]{iopart}

\usepackage{fancyhdr}
\fancypagestyle{plain}{%
\fancyhead{} 
}
\usepackage{geometry}                
\usepackage[parfill]{parskip}    
\usepackage[pdftex]{graphicx,color}
\usepackage{color}
\usepackage{epstopdf}
\usepackage{bm}
\usepackage{wrapfig}

\definecolor{dullmagenta}{rgb}{0.4,0,0.4}   
\usepackage{iopams}
\usepackage{bm}
\usepackage{graphicx}
\usepackage{amssymb}
\usepackage{hyperref}

\def\la{\label}
\def\be{\begin{equation}}
\def\beq{\begin{equation}}
\def\eeq{\end{equation}}
\def\ee{\end{equation}}
\def\bea{\begin{eqnarray}}
\def\eea{\end{eqnarray}}
\def\p{\partial}
\def\cprime{$'$}

\newcommand{\IM}{{\rm Im}\,}
\newcommand{\RE}{{\rm Re}\,}
\newcommand{\ii}{{{i}}}
\newcommand{\dd}{{{d}}}

\newcommand{\eq}[1]{(\ref{#1})}

\newcommand{\Pe}{Painlev\'e\;}

\begin{document}

\title{ Viscous shocks in Hele-Shaw flow and  Stokes phenomena of the Painlev\'e I transcendent }

\author{Seung-Yeop Lee$^1$, Razvan Teodorescu$^2$ and Paul Wiegmann$^3$}
\address{$^1$Mathematics 253-37, Caltech, Pasadena, CA 91125, USA}
\address{$^2$Mathematics Department, Univ. of South Florida, 4202 E. Fowler Ave, Tampa FL 33620, USA}
\address{$^3$The James Franck  Institute, University of Chicago, 5640 S. Ellis Ave, Chicago IL 60637, USA}
\eads{\mailto{duxlee@caltech.edu}, \mailto{razvan@usf.edu},
\mailto{wiegmann@uchicago.edu}}

\begin{abstract}

In Hele-Shaw flows at vanishing  surface tension, the boundary of a viscous fluid develops cusp-like singularities. In recent papers \cite{LTW1,LTW2}  we have  showed that   singularities trigger viscous shocks propagating through the viscous fluid. Here we show that the {\it weak solution of the Hele-Shaw problem} describing viscous shocks is equivalent to a semiclassical approximation of a special real solution of the Painlev\'e I equation. We argue that the Painlev\'e I equation provides an integrable deformation of the Hele-Shaw  problem which describes flow passing through singularities. In this interpretation shocks appear as Stokes level-lines of the \Pe linear problem.

\end{abstract}

\maketitle

\section{ Introduction}
Hele-Shaw flow describes a 2D viscous incompressible fluid with a free boundary at  low Reynolds numbers. The fluid is  driven to a drain by another, inviscid, incompressible liquid or a gas (see e.g., \cite{Couder}). Under this process the boundary of the fluid  evolves in an unstable manner, revealing singularities occurring at a finite time.  

The governing equations - Darcy's law -  can be easily derived from Navier-Stokes equations  under the assumption of a Poisseuille profile.  In units such that  the fluid density  and  hydraulic conductivity are set to one,  velocity of the fluid is proportional to the gradient of  pressure
\be \la{darcy}
\mathbf{ v} = - \mathbf{\nabla} p.
\ee
 In incompressible fluids, $\mathbf{\nabla}\cdot\mathbf{v}=0$ and pressure   is a harmonic function, $\Delta p=0$,  everywhere except at a drain (a marked point set at infinity), where the fluid disappears with  a  constant  flux $Q=\oint_\infty  \mathbf{ v}\times\dd \mathbf{ \ell}$ i.e. the amount of fluid removed in the time interval $\dd T$ is $Q\dd T$.

In the regime where the surface tension at the fluid boundary may be neglected, the case we consider, pressure is a constant $p|_\Gamma=0$ along the fluid boundary $\Gamma$.
\begin{figure}[h] 
   \centering
   \includegraphics[width=2.5in]{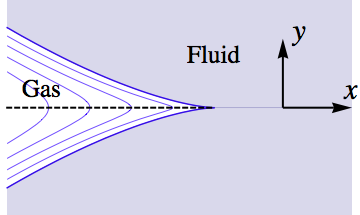}
   \caption{A finger growing in time (sequence of blue lines) toward the (2,3)-cusp singularity. The dashed line is the branch cut of the height function.}
   \label{finger}
\end{figure}

The Hele-Shaw problem is just one particular realization of a broad class of growth processes governed by the Darcy law. Darcy law states that  the normal velocity  of a line element of the boundary $d\ell$   is proportional to its  harmonic measure density. Harmonic measure is the probability of a Brownian excursion starting at a marked point (in this case it is a drain) to exit the domain through that boundary element. The probability  density of Brownian excursion is the Poisson excursion kernel, equal to the gradient $|\nabla  p|$  of a harmonic function $p$ with a source at a marked point and Dirichlet boundary condition. 

A well-known stochastic realization of the Darcy law is  DLA --  diffusion-limited aggregation \cite{DLA81}. It is realized through Brownian excursions of particles with a small size $\hbar$ emanating at a constant rate from a distant point, with absorbing boundary, such that the stopping-set cluster grows. The Darcy law emerges at vanishing   particle size,  $\hbar\to 0$.

Computer experiments with DLA  \cite{DLA81} show that a boundary, initially featureless, very quickly develops into  a branching  graph with a width controlled by  the size of one particle $\hbar$.  This signals that the limit $\hbar\to 0$ is impossible.

Darcy's law \eq{darcy} also indicates that Hele-Shaw flow tends to a microscale. Non-linear differential equation \eq{darcy} is ill-defined. A smooth initial boundary first evolves into a fingering pattern, then fingers   at a finite, {\emph {critical time}}  develop cusp singularities  \cite{bs84, Howison85,Hohlov-Howison94,TWZ}. At that point the differential form of the Darcy law stops making sense. This phenomenon constitutes the major problem in the field.

Unlike in fluid dynamics, DLA processes possess a dimensional parameter -- the  particle  size $\hbar$. It plays  the role of a minimal area, which regularizes singularities emerging in fluid dynamics.

Comparing Hele-Shaw flows to DLA suggests that, after a singularity is reached, the flow should feature shocks -  a graph of curved lines where pressure (and velocity) suffer discontinuities. Such solutions of ill-defined non-linear differential equations are known as {\it weak solutions} \cite{Evans}.

In Refrs. \cite{LTW1,LTW2} we developed the {\it weak solution} of Hele-Shaw problem, which  we believe to be    a solution of the problem of Hele-Shaw singularities. Within this solution,  a singularity triggers a branching graph of  shocks or cracks  propagating through the viscous fluid. Shocks at small Reynolds number is a new phenomena, not yet being observed experimentally. We refer them as {\it viscous shocks}. An emerging pattern of viscous shocks  is reminiscent   DLA patterns. 

The weak solution of the Hele-Shaw flow is based on an  integral form of the Darcy law.  This  form of the non-linear equation  suggests an interpretation of the Darcy law as an evolution of a real Riemann surface (a complex curve with reality conditions). A singularity signals that the evolving  Riemann surface changes its genus.

 A similar problem is known  in a different field - a slow dynamics of modulated  fast oscillatory non-linear waves (see e.g., \cite{Whitham}). If a non-linear wave features a  separation of scales, in the form of a slow modulation of fast oscillations, the slow part is described  Whitham equations \cite{Dubrovin-Novikov,Krichever}.   Conversely, the Whitham equations  can be interpreted as the evolution of Riemann surfaces\footnote{An observation that a   Hele-Shaw flow has a form of Whitham equations appeared  in Ref. \cite{MKWZ}.}

 In Refrs. \cite{LTW1,LTW2} we described hydrodynamics of  shocks initiated by the most generic
 cusp singularity,  referred to as a (2,3)-cusp.  In local Cartesian coordinates aligned with a cusp axis, the shape  of a critical finger  is  $y^2\sim x^3$, as in Fig.~\ref{finger}.

Right before  the occurrence of a (2,3)-cusp singularity,  a Riemann surface describing  Hele-Shaw flow is a sphere. Right after a singularity, the sphere transforms into  a torus (or elliptic complex curve). Exactly the same  transition appears  in a semiclassical approximation of  special, often called ``physical" solutions of the Painlev\'e I equation \cite{Gamba, Grinevich-Novikov}.

Appearance of the \Pe equation in Hele-Shaw dynamics is not accidental. In  this paper we clarify this relation. We show that the Hele-Shaw singular flow and  emerging shock pattern are directly related to  ``physical" solutions of Painlev\'e I equation: 
 \be\la{P}
  u_{ttt}-12u u_t=\hbar ,
 \ee
 where  the \Pe-function $u(t)$ is a real function for real values of the argument. 

\begin{figure}[hh] 
   \centering
   \includegraphics[width=2.5in]{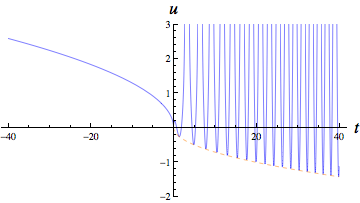} 
   \caption{Illustration of a physical \Pe-function by means of Whitham approximation: at $t<0$ the \Pe-function  is smooth $u\approx\sqrt{-\hbar t/6}$. It features fast oscillations  at $t>0$, but the change of the period and  amplitude remain smooth.  The (orange) dashed line is the envelop of minima $u(t_n)\approx 0.553594\sqrt{\hbar t_n/6}$. The picture captures only the asymptotes at large $|t|$, not  a crossover between smooth and oscillatory regimes.   }
   \label{Gamba}
\end{figure}

 As it is seen in  Fig.~\ref{Gamba}, the \Pe-function behaves drastically different at large negative times $t<0$ and large positive times $t>0$. At $t<0$ the \Pe-function is smooth. It turns to an oscillatory regime at $t>0$. 
 
 The \Pe-function may be linked to the position of the tip of the finger $e(T)$. The relation  changes  while  passing through the singularity:
  at the smooth phase (large negative $T$)   the tip progresses as $e(T)=-2u(t)<0, \;T\to-\infty$;  at large positive  $T$ (after the singularity) the tip 
 retreats, its position follows  a  slowly moving position of minima of the \Pe function $e(T)=u(t_n)<0$, where $t_n$  is a sequence of time, labeled by an integer $n$,   such that $\dot u(t_n)=0$. In both cases, the relation between a "slow"  time $T$  of the Hele-Shaw flow  and a "fast"  time  $t$  - an argument of the \Pe function  is 
  \be\la{T}
  T=\frac{\hbar t}{6},\; T<0;\quad  T=\frac{\hbar  t_n}{6},\; T>0.
  \ee

The Hele-Shaw flow is directly related  to the {\it semi-classical} ($\hbar\to 0$)  limit of the Painlev\'e ``wave-function"  $\Psi(X,t)$ -- solution of the linear  ordinary  differential equation  associated with Painlev\'e I equation:
\be\la{XY1}
\Big[\p_t^2-2u(t)\Big ]\Psi=X\Psi.
\ee
 The argument $X=x+iy$ (complex spectral parameter of the linear problem) plays a role of a complex coordinate  of the fluid. Shocks appear as {\it Stokes level lines} of the Painlev\'e wave-function. Hydrodynamic objects are obtained from the wave-function by averaging over fast time oscillations. In particular, pressure reads
 $$
 p(x,y)=-\lim_{\hbar\to 0}\RE\,\hbar\,\p_T\overline{\log \Psi(x+iy,t)}=-6\lim_{\hbar\to 0}\RE\,\p_t\log \Psi(x+iy,t_n),$$
 where overline means an average over fast time oscillations.  Alternatively one can  evaluate the wave-function at discrete  time moments \eq{T} .
 
%

We believe that the relation between the singular Hele-Shaw problem and the Painlev\'e I equation clarifies both entities. It further  explains the nature of weak solutions of Hele-Shaw problem and shocks in viscous fluids, and provides a   transparent, hydrodynamic interpretation of a subtle  semiclassical description of Painlev\'e I, which  otherwise is rather formal and complicated. In this paper we give a brief account of both sides of the story: the shock-wave solution of the Hele-Shaw flow and the Whitham averaging method  for the physical solution of Painlev\'e I, even though elements for each of these themes taken separately can be found in the literature. We have found that  approach to \Pe-equation as an isomonodromic deformation developed in  Refrs. \cite{Jimbo1,Kapaev1989,Takei,Kapaev2001,Kapaev2004,Fokas2} provides the closest connection between \Pe-equation to  Hele-Shaw hydrodynamics.

The appearance of the Painlev\'e equation in the Hele-Shaw problem is more than a formal analogy. \Pe equation gives an integral Hamiltonian  regularization of the Hele-Shaw singularities anticipated in Ref. \cite{TWZ}.

Contrary to the Darcy law, the Painlev\'e equation (like a DLA process), has a dimensional parameter  $\hbar$. The meaning of this parameter can be understood as follows. The physical solution of \Pe I-equation is selected by the asymptote $u\to  +\sqrt {-T}>0$ at large negative  times  $T\to-\infty$. At large  positive times $T>0$,  this solution features fast oscillations  with a period of the order of $ \hbar $. The period of oscillations  also  changes in time, but slowly,  such that the period  in units of $\hbar$  is a regular function  at  $\hbar\to 0$.  As such, $u(T)$ does not have a limit at $\hbar\to 0$, but its period  (in units of $\hbar$) does.  The Hele-Shaw flow arises from the Painlev\'e equation in the limit  when a dimensional parameter $\hbar$ vanishes.  It features singularities and discontinuities while the Painlev\'e equation at $\hbar\neq 0$ does not.  At a non-vanishing $\hbar$  shocks are regularized. They appear  as spatial narrow regions  of width $\sim \hbar^{3/2}$ where pressure and velocity exhibit finite,  though sharp, gradients.  In this sense, the Painlev\'e equation provides a regularization of the Hele-Shaw flow, similar to a regularization that provides  DLA. 

Emergence of  singularities in  hydrodynamics is typically caused by an invalid approximation  made in an original  physical problem. Any physical problem of fluid dynamics always  carries a dimensional parameter, which regularizes singularities at a microscale. As a result,  physical processes are smooth, but may exhibit large gradients in certain regions of time and space. When the microscale parameter $\hbar$ is naively set to zero in the equations, they may  become ill-defined and do not have physical solutions in the regime when  gradients are large. However, if the parameter is set to zero in the solution (rather then in the equation), solutions retain their meaning but may become discontinuous, exhibiting shocks --  {\it weak solutions} \cite{Evans}. This is what happens in the  Hele-Shaw problem.

Painlev\'e equation itself is an example of these phenomena. Setting $\hbar=0$ in the equation reduces it to $-2uu_T=1$, whose solution $\sqrt{-T}$ makes sense (i.e., remains real) only at $T<0$.  Although at $T>0$ the \Pe-function has no formal  $\hbar\to 0$ limit, its average  over the period of fast oscillations or  a  motion of  minima of \Pe-function does  (as it is seen on the Fig.~\cite{Gamba}).   We will show that this  slow motion is equivalent  to the weak solution of the Hele-Shaw problem obtained in Refrs. \cite{LTW1,LTW2}.

In many important cases, weak solutions show a great degree   of universality -- a large class of regularized problems lead to the same weak solution. It is always interesting to determine   regularizations that are ``minimal" in some sense. In this paper, we suggest that the \Pe equation provides  the  ``minimal" regularization for generic  singularities of the  Hele-Shaw problem. It is  the only known integrable regularization.

The nature of the physical solution of the Painlev\'e equation suggests a physical interpretation of the dimensional parameter $\hbar$. The solution suggests that, under the regularization, the inviscid fluid consists of particles with a minimal area $\hbar$. Oscillations reflects a discreetness of particles and a "corpuscular" nature of the problem.  This interpretation links the Painlev\'e I regularization of the Hele-Shaw flow with the DLA process where particles are essentially discrete, and provides an analytical basis to study the latter.

In the next section we briefly review the  Hele-Shaw critical fingers in terms of evolution of the  real complex curve. It provides a necessary  language to link the flow with the Painlev\'e wave-function described in the sequel.  Our approach is based on isomonodromy property of \Pe-equation. Our approach  is somewhat different that in the Refrs. \cite{Kapaev1989, Kapaev2001,Kapaev2004, Fokas2}. In particular we give a new simple derivation of the expression for the phase  \eq{phase} and the next leading oscillatory  corrections \eq{barO},\eq{barH} in Sec. ref{MKB}  .

A comment on  related studies  is in order.  In Refrs. \cite{ABZW,ABTWZ} Hele-Shaw flow has been linked to normal random matrix ensembles, and through them to certain asymptotes of bi-orthogonal polynomials. Random matrix ensembles  provide an equivalent regularization scheme of Hele-Shaw flow.  \Pe equation can be formally derived from this approach \cite{ABTWZ, T}.  In  a similar manner \Pe equation  appears  in studies of Hermitian random matrix ensembles \cite{David} and related to their asymptotes of orthogonal polynomials \cite{Fokas, Belgians}. From this perspective a singular Hele-Shaw flow is linked to a distribution of zeros of orthogonal and bi-orthogonal  polynomials and eigenvalues of random matrices.  In this paper we do not elaborate this approach in order to reduce the size of the paper.

We also comment  that an idea that near-cusp structure should be followed  by a crack propagation into the fluid  domain has been suggested in earlier works \cite{crack1,crack2}. The mechanism discussed in these papers seems  different from that in \cite{LTW1,LTW2}.

\section{Singularity and viscous shocks}\la{S}
 \subsection{Evolving  complex curve}

   The flow near a singular  finger tip does not depend on details of a  boundary away from the tip of the finger. It is sufficient to consider a finger with a symmetry axis.   In Cartesian coordinates
 aligned with a finger axis (Fig.~\ref{finger}), a  finger  is given by  a graph $\pm y(x)$.
We recall a description of a viscous finger in terms of the \emph{height} function $Y(X,T)$ \cite{TWZ}. It is defined as an analytic function on a finite part of the  complex plane  $X=x+iy$ cut along the symmetry axis of the finger, such that   the boundary values are   real and equal to the graph of the finger $Y(X)_{X=x+\pm\ii 0}=\pm y(x,T)$.
Darcy law can be  expressed through the height function \cite{TWZ}:
   \be\la{darcy2}
  \partial_T Y= -\p_X\phi,
  \ee
  where the analytic function $\phi(X)=\psi+\ii p$ is a complex potential of the flow,  $\psi$ is a  stream function,  and $p$ is the pressure.

The height function $Y(X,T)$ defines a complex curve (or Riemann surface) evolving in time. Since at real $X=x$ the height $Y=\pm y(x)$ is real, the curve is real: $\overline{Y(X)}=Y(\overline X)$.

In Ref.\cite{TWZ} it has been shown that in general the complex curve is hyperelliptic, i.e.,  $Y^2(X)$ is a polynomial  of an odd degree.  Its degree is preserved while the coefficients evolve in time.

 The flow potential $\phi(X)$ has an important interpretation. Darcy law states that $\phi$ is a local parameter of the curve. This means the following: the potential $\phi(X)$ is an  analytic function. Locally it can be inverted, so that $X$, and other physical quantities, say, the height function,  become   functions of $\phi$. Then Darcy law \eq{darcy2} states that functions $\left(Y(\phi),\, X(\phi)\right)$ treated as functions of the potential, are  regular and single-valued.  In the case of critical flow,  when $(Y, \,X)$ is hyperelliptic, the variable $\phi$ covers the physical plane twice: $ \phi$ and $-\phi$ correspond to the same $X$, such that the double-valued function $Y(X)$ becomes a single-valued function $Y(\phi)$  on the double covering.

The fact that the  potential is a local parameter of the curve  is the essence of the Darcy law.  The law can be formulated entirely  in terms of the curve: the curve $(Y,\, X)$, with a local parameter $\phi$, moves in time such that the differential $\ii\dd\Omega=Y\,\dd X+\phi\, \dd T$ is closed. In this form Darcy law becomes identical to Whitham equations
describing slow modulation of fast oscillating non-linear waves \cite{Whitham, Flaschka-McLaughlin, Krichever}.
This fact was recognized in \cite{MKWZ}.

A consequence is Polubarinova-Kochina 
 equation \cite{Galin,PK} (which in modern literature  is called ``semiclassical string" equation \cite{DiFrancesco} ).  Treating $Y$ and $X$ as  functions of the complex potential $\phi$, the Darcy law  \eq{darcy3} reads
\be\la{33}
\{X,\,Y\}=1,
\ee
where the brackets are defined via $\{X,\,Y\}=\p_T X \p_\phi Y-\p_\phi X\p_T Y$ and  time derivative is taken at fixed $\phi$.

The flow is best described by  a generating function -- a  central object of this study:
\be\la{O}\Omega(X)= -\ii \int_e^X Y\dd X,
\ee
 where the integration starts at the tip of the finger $e$ and goes through the fluid.
  Then the  Darcy law reads
 \be\la{darcy3}
 \p_T \Omega = \ii \phi. 
 \ee

\subsection{Integral form of Darcy law}

Darcy law can be equivalently expressed by integration of the differential $\dd\Omega$ along various paths  in the fluid.
Let us  integrate (\ref{darcy2})  over a closed path  $B$ in the fluid:
 \bea \la{im}  \frac{\dd \,\,}{\dd T}{\rm{Im}}\, \oint_{B}   \dd \Omega
&=&\!\!\oint _B \mathbf{v\times d\ell}
= \oint_B \dd\psi,\\
\la{re}  \frac{\dd \,\,}{\dd T} {\rm{Re}}\,
\oint_{ B}    \dd \Omega &=& - \oint _B  \,\mathbf{
v\cdot d\ell}
=  - \oint_B \dd p.
\eea

The imaginary part measures a flux of fluid  through the closed path. If the path goes around a drain (infinity),  the integral $ {\rm{Im}}\,\oint_{ B}    \dd \Omega$ counts  a  mass of fluid drained up to time $T$. It is  $QT$.  In a canonical anti-clockwise orientation this reads
\be\la{11}\ii  \frac{\dd}{\dd T} \oint_\infty \dd\Omega=Q>0.
\ee

 The real part measures circulation along the path.  It follows from the differential form of the Darcy law (\ref{darcy}), valid before a singularity, that the fluid is irrotational $ \mathbf{\nabla\times v}=0$ and therefore
$\frac{\dd}{\dd t} {\rm{Re}}\,\oint_{ B}    \dd \Omega=0$.  Therefore ${\rm{Re}}\,\oint_{ B}    \dd \Omega=0$  is  a constant.  Provided that the constant does not depend of the choice of the path,  the constant can be determined by taking the integral around a drain where $ \oint_\infty \dd\Omega$  is purely imaginary as is in (\ref{11}). Thus 
${\rm{Re}}\,\oint_{ B}   \dd \Omega=0$ on any contour in the fluid.

If there are shocks, the height function and $\dd\Omega$ jump through a shock. The condition ${\rm{Re}}\,\oint_{ B}   \dd \Omega=0$  for a path crossing a shock must be understood as a sum of the integral over a path in the fluid plus an oriented jump of $\dd\Omega$ at points where the path crosses a shock. A proper way to write this condition is to extend the integration over the Riemann surface $(Y,\,X)$ where the  height function and the differential $\dd\Omega$  are smooth. The physical plane where the hight function  suffers  a discontinuity consists onn patches of  different sheets of the curve.  Then the above condition reads
\be\la{22}  {\rm{Re}}\,\oint_{ B}   \dd \Omega=0,\quad \rm{any\;cycle  \;on \;the\;curve.}
\ee

Curves with  condition (\ref{22}) are rather restrictive. We call them Krichever-Boutroux curves. They appeared in seemingly different, but related subjects: random matrix theory \cite{David}, asymptotes of orthogonal polynomials \cite{Bertola} and  Painlev\'e  transcendents \cite{Boutroux, Krichever1, Kapaev-Kitaev}.

Eqs. (\ref{11},\ref{22}) combined give a compact and complete  formulation of the Hele-Shaw problem:
find an evolution of  a real Krichever-Boutoux curve (condition  (\ref{11})), of a given degree,  with respect to its residue (time $T$)  at a marked point (condition  (\ref{11})). 

Contrary to the differential form of the  Darcy law \eq{darcy},  Eqs. (\ref{11},\ref{22})   extend beyond singularities. They  admit a weak solution with shocks. Shocks are curved lines where patches  of the complex curve covering the physical plane are joined together. At these lines, 
the height function  jumps. Condition (\ref{22}) being specified for a contour going along both sides   of a shock reads
\be\la{66}\rm{Re}\,\rm{disc}\,\Omega|_{\rm{shocks}}=0.\ee
These conditions determine position and motion of shocks.
 These are the  results of Refs. \cite{LTW1,LTW2}.

In the next two sections we demonstrate how  these conditions determine the  most generic singularity  and also physical solution of the  Painlev\'e I equation.

 \subsection{(2,3)-singularity and a self-similar elliptic curve}
Painlev\'e I equation is related to the most generic (and the most important) singularity -- the (2,3) cusp, where locally  the graph of the finger is $y^2\sim x^3$.
%
 In this case a  finger  is especially simple \cite{TWZ}:  $Y^2$ is a polynomial of the third degree - an elliptic curve having one branch point at infinity.
Fixing a scale and the origin, the curve  reads:
 \be\la{e}
- Y^2=4X^3-g_2X-g_3=
 4(X-e)(X-e_2)(X-e_1),
  \ee
  where $g_2$ and $g_3$ are real time dependent coefficients.
 One of  the branch points of the curve (\ref{e}) may always  be chosen real. We denote it as $e$. It is the tip of a finger, where $y(e)=0$. The other two  are conjugated $e_1=\bar e_2$ (condition (\ref{22}) excludes a possibility of real
 $e_1$ and $e_2$ - in this case $\rm{disc}\,\dd\Omega$ through the branch cut connecting $e_1$ and $e_2 $ is real  and does not change sign, so that   $\RE\oint_{e_1,e_2}\dd\Omega$   can not vanish).  This is a real elliptic curve with a negative discriminant $g_2^3-27g_3^2<0$.  The coefficient $g_2$ is determined by the drain rate. Eq. (\ref{11}) gives  $Q \sim  \dot g_2$. We set the rate $Q$ such that  
   \be\la{G2}g_2=-12T.
   \ee
    In this setting the time   counts from a cusp event $-Y^2=4X^3$  occurring at a critical time $T=0$. A real coefficient $g_3$ is determined by the condition (\ref{22}).

This is the simplest curve arising in singular flow. Its distinctive property is self-similarity.  As it follows from \eq{11},\eq{22}
\be\la{s}
Y(X,T)=|T|^{3/4}Y\left(|T|^{-1/2}X,1\right),
\ee
where $Y(X, 1)$ is a unique universal function with no free parameters.

The scaling property  alone is sufficient to
 express  $\Omega$ through the height function and potential:
 \bea \la{phi}
 &&\Omega(X)=-{\ii}\int_{e}^X Y\,dX=-\frac{2\rm i}{5}(XY-2T\phi).\la{Omega}
\eea

 We briefly describe this curve following Refs. \cite{LTW1,LTW2}.

\subsection{Smooth flow and a degenerate curve}
At $T<0$, before  the singularity,  the flow is smooth, there are no discontinuities.  This follows also from the condition \eq{22}.  At $g_2>0$ only degenerate elliptic curve whose discriminant  vanishes,  $g_2^3=27g_3^2$.  The  only unknown coefficient  is thus determined, $g_3=-8(-T)^{3/2}$.  The two branch points 
 coincide to a real  double point:  $e_1=e_2=+\sqrt{-T}>0$.   Then $e=-2e_1=-2\sqrt{-T}$. The degenerate curve 
  \be\la{133}
Y^2=-4 \left(X-e\right)\left(X+\frac e2\right)^2,\quad e=-2\sqrt{-T},\quad T<0
\ee
represents a torus with one pinched cycle.
The real period of the curve is infinite, the imaginary half-period is $\int_{e}^{-\infty}\frac{dX}{Y}=\pi\ii \sqrt{12}(-T)^{-1/4}$. 
This is a general situation - a continuous flow corresponds to a degenerate curve where all cycles but one are pinched   \cite{TWZ} (the number of non-degenerate cycles is equal to the connectivity of the fluid boundary \cite{MKWZ}). Condition that all cycles   located in the bulk of the fluid are pinched follows from incompressibility of the fluid. Only in this case  the fluid potential $\phi$ is analytic.

Alternatively, we notice two holomorphic functions $Y(\phi)$ and $X(\phi)$ with  an  asymptote $Y^2\sim X^3$ at large $X$  are polynomials of the third and second degree, respectively.  The Darcy law in the form of (\ref{33})  determines the  polynomials 
\be\la{44}
X=-\left(\frac \phi 6\right)^2+e(T),\quad Y=2\left(\frac\phi 6\right)^3-3e(T)\left(\frac\phi 6\right),
\ee
and yields to a differential  equation for time dependence of the coefficients
\be \la{ee}
e \p_T e= -2,
\ee
which prompts \eq{133}.  Real solution exists only at $T<0$.

\subsubsection {Beyond singularity: Genus transition.}\la{G}

A finger becomes a cusp when the  branch point $e$ and the  double point $-\frac e2=e_1=e_2$ merge to a triple point $e=e_1=e_2=0$
\be\la{77}
Y^2=-4X^3,\quad T=0.
\ee
After the critical time $g_2=-{12}T<0$, the curve is not degenerate. The double point splits into two branch points $e_1\neq e_2=\bar e_1$. No  smooth solution is possible. The branch points  appear in the fluid as endpoints of shocks.  Fig.~\ref{physicalplane} illustrates the genus transition. A pinched  cycle of the torus now opens.   

The  condition (\ref{22})  requires that the integral over a new cycle  be  purely imaginary
\be\la{BBB}
\Omega (e_1)=\int_0^{e_1}\sqrt{4X^3-g_2X-g_3}\,\dd X=\mbox{imaginary}.
\ee
Exactly the same equation has been appeared in works of P. Boutroux \cite{Boutroux} on \Pe I equation. We will refer it as Boutroux condition.
This equation can be simplified: since $Y(e_1)=0$, scaling properties \eq{Omega} yield that at  the end of a shock $e_1$, pressure $\phi(e_1)=
0$ also vanishes.
The latter  uniquely determines the only unknown parameter $g_3(T)$ of the curve.
%

\begin{figure}[htbp]
\begin{center}
\includegraphics[width=6cm]{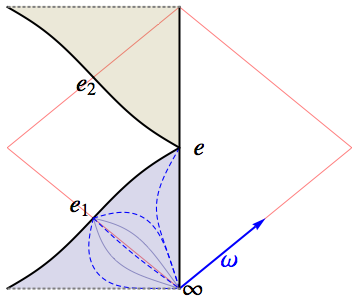}
\caption{ 
Stokes-level   (dashed-blue) lines $\RE\Omega=0$ and Stokes lines  (solid black) $\IM\Omega=0$   on the fundamental domain of the double covering of the physical plane.  The blue-shaded region maps to the upper half-plane in the physical plane.  The other  shaded region maps to the lower half-plane.}
\label{rhombus}
\end{center}
\end{figure}
Let us parametrize the curve $Y(X)$ by a uniformizing coordinate $\lambda$ as
\be\la{20}
X=\wp(\lambda|\,\omega(T),\bar\omega(T)), \quad
Y={i}\wp'(\lambda|\,\omega(T),\bar\omega(T))
\ee
and compute potential  of the flow from the Darcy's law and scaling property (\ref{s})
\be\label{phi1}
-i\p_T\Omega=\phi(X)=6{\ii}\left(\zeta(\lambda)+\frac{g_3}{8T}\lambda\right).
\ee
Here $\wp$ and $\zeta$ are  the elliptic functions of Weierstra{\ss}, whose complex conjugated half-periods $\omega(T)=T^{-1/4}\omega(1)$ and $\overline{\omega(T)}$ are yet to be determined. The rhombus-shaped  fundamental domain is depicted in  Fig.~\ref{rhombus}.

Conditions that pressure vanishes at $e_1$ and formula \eq{phi1}  give the defining equation:
 \begin{equation}\label{phib}
 \frac{3 g_3}{2g_2}=\frac{\eta+\bar\eta}{\omega+\bar\omega},
\end{equation}
where $\omega+\bar\omega$ is a real period of the curve and  $\eta=\zeta(\omega)$.

The solution of this equation  is   found in \cite{Gamba}
\bea\la{21}
g_3&=&|T|^{3/2}
\left\{
\begin{array}{lr}
-8,&\quad T<0,\\
-7.321762431\dots,&\quad T>0.
\end{array}
\right.\\
\la{periods}
 \omega&=&T^{-1/4} (0.7426778188 + i \,0.6070379047).
\eea
The  genus transition is characterized by an  abrupt change of $g_3$.
The modular invariant of the non-degenerate  curve $ \frac{g_2^3}{g_2^3-27g_3^2}=6.15870507\dots$ is a unique transcendental number.

 \subsection{Motion of the finger}
 The finger is described by the real section of the curve $Y,X$, i.e., by the graph $y(x)=Y(X)$ where $X=x\leq e(T)$. As it follows from the (\ref{133}) a finger grows and its tips moves to the right as its approaches toward a cusp singularity. After the singularity, the finger retreats, moves to the left and triggers two  shocks  growing to the right.  The tip of the finger gives the position of the branch cut extended toward $-\infty$. The scaling property shows that tip of the finger moves as $\sqrt{|T|}$ but with an abrupt change of the coefficient \cite{LTW1,Gamba}
 \begin{equation}\la{e1}e(T)=
\left\{
\begin{array}{lr}
-2\sqrt{-T},&\quad T<0,\\
-0.553594\sqrt{T},&\quad T>0.
\end{array}
\right.
\end{equation}
In the next section we show that  the envelope of the \Pe function behaves in a similar manner
 \begin{equation}
 u(T)=
\left\{
\begin{array}{lr}
-\frac 12e(T)
=\sqrt{-T},&\quad T<0,\\
\;\;\;\;\;e(T)=-0.553594\sqrt{T},&\quad T>0.
\end{array}
\right.
\end{equation}

\subsection{Shocks and Stokes level-lines lines.}\la{SS}

\begin{figure}[h]
\begin{center}
 \includegraphics[width=2.9in]{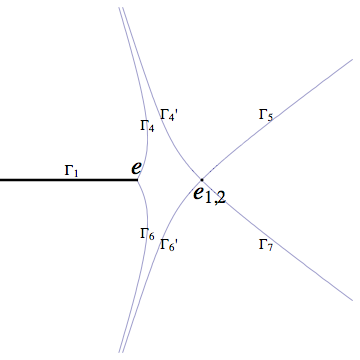} 
\includegraphics[width=2.9in]{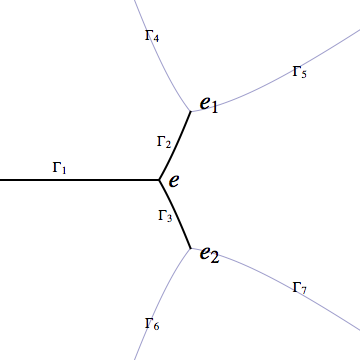}
\caption{ Geometry of Stokes-level  lines  (black solid lines) before (left panel)  and after (right panel) transition.  Before the transition four level lines emanate from the double point and end at infinity, neither one  is a shock.  After the transition the double point splits into two branch points.  Condition  ${\rm Re}\,\Omega >0$ on the remote part of the finger ${\rm arg}\,X=\pi,\;|X|\to \infty$  uniquely determines configuration of shocks $\Gamma_3, \Gamma_2$. They are always located in the finite part of the plane.}

\label{physicalplane}
\end{center}
\end{figure}

Shocks are  lines of discontinuity of the  generating function $\Omega$, the height function $Y$ and the potential $\phi$.  They are selected by the condition \eq{66}, which in the case of simple branch points says that shocks are a subset of zero level lines  ${\rm Re}\,\Omega=0$. We refer to the lines   ${\rm Re}\,\Omega=0$ as  Stokes level-lines.  There are a total of seven Stokes level-lines connected at  branch points. They are transcendental, computed numerically, and depicted on the right panel in Fig.~\ref{physicalplane}, denoted by $\Gamma_1,..., \Gamma_7$.

Among the seven Stokes level-lines, only two  $\Gamma_3, \Gamma_2$ are shocks.
 Both sides of the line $\Gamma_1$ are boundaries of the finger. 

Selection of shocks works as follows. We assume that a  remote part of the finger  (${\rm arg \,X}=\pi,\;|X|\to\infty$) is not affected by the transition.  Therefore at large $X$ the height function $Y(x)$ jumps only on the finger, i.e., on the negative part of real axis. This excludes Stokes level-lines lines $\Gamma_{4,5,6,7}$   ending at infinity.
Thus $Y$ jumps only on  finite Stokes-level  lines $\Gamma_{2,3}$. On a shock, $Y$ (as a complex vector) is orthogonal to the shock. This uniquely determines the shape of shocks (Fig.~\ref{physicalplane}).

The left panel in Fig.~\ref{physicalplane}  represents Stokes level-lines before the transition $T<0$. There are 9 lines. Four Stokes level-lines $\Gamma_4',\Gamma_5,\Gamma_7,\Gamma_6'$ emanate from the double point. All of them end at infinity. Neither of the Stokes level lines is a shock.

Having in mind the  connection to the Painlev\'e wave-function discussed in the next section, it is useful
to translate the selection rule in terms of the generating  function: the signs of ${\rm Re}\,\Omega$
 are opposite on both sides of $\Gamma_{4,5,6,7}$, and signs are positive on both sides of $\Gamma_1,\Gamma_2,\Gamma_3$.  This is a general situation: Stokes-level  lines  ${\rm Re}\,\Omega=0$ are shocks if
 ${\rm Re}\,\Omega$  
  has the same sign (positive)  in the proximity of both sides of a shock \cite{LTW1, LTW2}
\be\la{ad}{\rm Re}\,\Omega|_{X\to\rm{shocks}}>0,\quad {\rm Re}\,\Omega|_{X\in\rm{shocks}}=0.\ee
In the next section we show how the same objects appear in the context of non-linear Stokes phenomenon for the  Painlev\'e  I  equation.

\section{Hele-Shaw flow and physical solution of Painlev\'e I }

\subsection{Linear problem for Painlev\'e I}
Painlev\'e  I equation arises as a result of compatibility of two linear differential equations satisfied by the``wave-function"  $\Psi(X,t)$ -- a central object of this study.

Let  $\hat X$ and $\hat Y$, defined through
\be\la{XY}
\hat X=\p_t^2-2u(t),\quad -i\hat Y=-2\p_t^3+6u \p_t+3u_t
\ee
be two Hermitian differential operators.  Then the linear problem  reads
\be\la{L}
\hat X\Psi=X\Psi,\quad \hat Y\Psi=\ii\hbar \p_X\Psi.
\ee

The two equations are compatible only if $u(t)$ obeys Painlev\'e I equation
$
u_{tt}-6u^2=\hbar t.
$
Only then
\be\la{XY2}
[\hat X,\,\hat Y]=-i \hbar,
\ee
as it follows from \eq{L}.

\subsection{ \Pe equation as a quantized Hele-Shaw flow}\la{A0}
Equation \eq{XY1} and \eq{XY} are to be compared with \eq{33} and \eq{44} describing a smooth Hele-Shaw flow. One identifies the operators $\hat X,\,\hat Y,\,\hat\phi,\, u$ with their classical counterparts
$X,\,Y,\,\phi,\, -e$, where  we define $\hat\phi:=-\ii\hbar\p_T=-6i \p_t$.  From this point of view Painlev\'e equation is seen as a quantization of Hele-Shaw problem, when the brackets  in  \eq{33} are replaced by the commutator in \eq{XY1}:\;$\{\,,\,\}\to \frac i \hbar[\,,\,]$.
 The quantization introduces a new parameter $\hbar$ which regularizes singularities of the flow. It will be sent to zero when a solution is found. 
This point of view has been suggested in \cite{TWZ}.

As we discussed in the introduction, the limit $\hbar\to 0$ exists only before a flow reached a singularity. After this point solution oscillates and formally does not have a limit. However its average over the period of oscillations does.

Quantization suggests a regularizing scheme for the Hele-Shaw flow: observables in the classical flow are diagonal matrix elements of the corresponding operators averaged over the period of oscillations. One may consider  the height function, flow potential,  and velocity of a regularized flow defined as an expectation  value of corresponding operators averaged over the period
\bea
&\langle\hat Y\rangle:=\overline{\Psi^{-1}\left(\ii\hbar \p_X\right)\Psi},\nonumber\\
&\langle \phi\rangle:=\overline{ \Psi^{-1}\left(-\ii\hbar\p_T\right)\Psi,\la{Y}}\\
&\langle v\rangle:=\overline{\Psi^{-1} \left(\ii\hbar\p_T\p_X\right)\Psi},\nonumber
\eea
where $\overline{\cal O}=({ \rm period})^{-1}\oint {\cal O} dt$ is an average over the period. These quantities have a regular (asymptotic) expansion in $\sqrt\hbar$.



In the remaining part of the paper we will show that a semiclassical limit of the  physical Painlev\'e wave-function describes the Hele-Shaw flow before and after a singularity. This statement is summarized as:   the generating function $\Omega=-\ii\int^XYdX$, \eq{O}  of the flow is the Hamilton-Jacobi action of the semiclassical approximation 
\be\la{OO}\Omega=\lim_{\hbar\to 0}\hbar\,\overline{\log \Psi(X,t)}.\ee

\subsection{Physical wave-function and asymptotes at infinity}
The general solution of  the \Pe equation is a  two-parametric family. Among them we must select a physical solution which we corresponds to   the Hele-Shaw  flow by means of \eq{OO}.    A central physical principle we have already used as a guideline is:   flow around a finger  does not affect the flow at infinity.  The height function and the potential   are   holomorphic functions around infinity (the drain)  $\phi \sim -6\ii X^{1/2}, \, Y\sim  2\ii X^{3/2}$  at $X\to\infty,\; |\mbox{Arg}X|<\pi$, as it follows from \eq{44}.  In terms of the wave function, these conditions  read: the asymptote of the  physical wave-function at infinity is an analytic function in the plane cut along the finger  $\mbox{Arg}X=\pi$
\be\la{V}
\hbar\log\Psi|_{X\to\infty}=\Omega|_{X\to\infty}\to  \frac{4}{5}X^{5/2}-t\,X^{1/2}, \quad  |\mbox{Arg}|X<\pi.
\ee
Also, we must impose two additional  conditions. One is reality
\be\la{Re}
\overline{\Psi(X)}=\Psi(\overline  X).
\ee
The second is that at $t=6T/\hbar \to-\infty$ the Boutroux condition \eq{22} holds
\be\la{KB}
\lim_{\hbar\to\infty}\hbar\,\RE\oint \dd\log\Psi=0, \quad  t\to -\infty.
\ee
As a consequence
\be\la{A1}
u(t)\to -\frac{e(T)}{2}=\sqrt{-T}=\sqrt{-\frac{\hbar t}{6}} >0, \quad t\to -\infty
\ee
where $e(T)$ is a coordinate of the  tip of the finger and $ e(T)/2$ is a position of the double point.

These simple requirements determine both the physical \Pe function and  the physical \Pe wave-function.  

\subsection{Viscous shocks and Stokes phenomena of Painlev\'e transcendent}
 The Painlev\'e wave-function is an entire function. It is differentiable everywhere in $X$. However, its semiclassical limit is not. The gradient $\Psi^{-1}\hbar\,\p_X\Psi$ is of order one in most part of the plane, except some lines where it changes abruptly on a scale vanishing with $\hbar$ across these lines.
At $\hbar\to 0$ sharp gradients become discontinuities -  seen as shocks in viscous  flow. The origin of these lines is simple: they go through the locus of zeroes of the wave-functions. Zeros are separated by $\hbar $ (at the end of the line the separation is $\hbar^{2/3}$), and  at $\hbar\to 0$ accumulate to a line distribution. Gradients through lines of zeroes are large.

Discontinuity of the semiclassical approximation of analytic functions is known as  {\it Stokes phenomenon}. The locus of zeroes is closely related to Stokes phenomena. Shocks occur on some Stokes level-lines  selected by the condition (\ref{ad}) and described in more details in the sequal.

 The governing conditions \eq{Re} and \eq{V} determine  the Stokes multipliers and semiclassical wave-function.
 
\section{The physical solution of Painlev\'e I transcendent}

Among many solutions of Painlev\'e I equation 
\be\la{P1}
 u_{tt}-6u^2=\hbar t
 \ee
 called Painlev\'e functions (see e.g.,\cite{Holmes, Kapaev1989}),  we look for the solution which  at $t\to -\infty $ 
 corresponds to a smooth  Hele-Shaw flow as in \eq{A1}. This correspondence also establishes a relation \eq{T} between the time of the Hele-Shaw flow $T$ and the argument of \Pe-function $t$ a relation between times . The correspondence selects a one parametric family of \Pe-functions. Following    \cite{Grinevich-Novikov} we call this solution ``physical" (it is also known as  ``separatrix" \cite{Kapaev1989} ).
 
This solution is scale invariant under the rescaling
  \be\la{scale}u\to\delta^{-2/5} u,\, t\to \delta^{-4/5}t ,\,u_t\to\delta^{-3/5} u_t, \, \hbar\to\delta\hbar.\ee 
 Below we review some major facts about asymptotes of the physical \Pe-functions and  derive most of them. 
 
 \subsection{Smooth regime $t\to -\infty$}
 The sign of the  asymptote $u|_{t\to-\infty}>0$ fixes the coefficients of the  asymptotic series
 \be\la{u}
 u(t)-\frac{|e(T)|}{2}= -\frac{|e(T)|}{2}\sum_{k=1} a_k(-t )^{-5k/2},\quad t\to -\infty,
 \ee
 where $e(T)$ is given by \eq{A1} and  $t=6T/\hbar$.
 
 The coefficients $a_k$ can be  computed recursively, $a_1=\frac{1}{48}$. They have a topological meaning as intersection numbers of curves on Riemann surfaces \cite{Witten}, and grow as $a_k\sim \frac{1}{\pi^{3/2}}\left(\frac 65\right)^{1/2}\left(\frac{2^5}{5}\sqrt 6\right)^{-k} \left(2k\right)!$ \cite{Takei,Joshi-Kitaev,Kapaev2004}.

The asymptotic series  alone does not fix the \Pe function. A one-parametric family  of solutions  shares  the same series, but  differs by the amplitude of an exponentially small  correction:
  consider  a difference between two different solutions sharing the same power series $u_1$ and $u_2$. It obeys the equation $(u_1-u_2)_{tt}=6(u_1+u_2)(u_1-u_2)$. Replace $u_1+u_2$ by the power series, and solve the linear equation with respect to $u_1-u_2$.   In the first approximation we obtain $(u_1-u_2)_{tt}\approx-12 e(T)(u_1-u_2)$. The large (negative)  time asymptote is controlled by the  WKB  approximation:   
  $$  u_1-u_2\sim \,e(T)\frac{e^{-\frac 45\sqrt{-6e(T)}(-t)}
}{\sqrt{2\pi\sqrt{-6e(T)}(-t)}} $$  
This formula   is valid in the sector $|{\rm Arg}\, t-\pi|\leq \frac{3\pi}{5}$ if $(-t) $ here and in \eq{A1}   is understood as $e^{-\ii\pi} t$ on the complex plane of $t$ cut along the negative real semi-axis. 
  
  Along Stokes level lines  $\mbox{Arg}\, t=\pi\mp\frac{2\pi}{5}$ in the complex plane of $t$ at $|t|\to\infty$ the exponential term  becomes of order one. There   the  asymptote of the physical  \Pe-function reads  \cite{Kapaev1989,Kapaev-Kitaev,Kapaev2004,Takei}
  \be\la{exp1}
  u(t)+
 \frac{e(T)}{2}\approx -\frac{e(T)}{2}\frac{C\mp\frac\ii 2}{2} \frac{e^{-\frac 45\sqrt{-6e(T)}(-t)}
}{\sqrt{2\pi\sqrt{-6e(T)}(-t)}},\quad \mbox{Arg}\, t=\pi\mp\frac{2\pi}{5}  \ee
where a real parameter $C$ uniquely characterizes the solution (these formulas are written in a form  emphasizing slow and fast time dependence   $e(T)=-2(e^{i\pi } \hbar t/6)^{1/2}$).  The  amplitude $C\pm\ii/2$ jumps through the negative part of the real axis. The jump is imaginary  and is the same for all physical solutions. The jump is linked to the large order behavior of the coefficients of the asymptotic power series   (\ref{u}) and can be obtained through  the Borel transformation technique \cite{Takei}.

\subsection{Oscillatory regime:  $t\to+\infty$}\la{O1}
As $t\to+\infty$ the family of solutions  features fast oscillations with a period  of  order $\hbar$,  and has infinite number of double poles clustering at infinity  as shown in Fig.\ref{Gamba} (there are other double poles in the complex plane of time).  Semiclassical description of the solution at $t\to\infty$ is based on a separation of scales between fast oscillations and the slow-varying period. Separation of scales is   controlled by the parameter $\hbar$, or large $T$.

At positive large time, the physical  Painlev\'e functions are asymptotically    Weierstrass elliptic function
\be\la{u1}
u(t)\approx \wp\left(\frac 4 5t-\theta(T)|\,\omega(T),\bar\omega(T)\right),\quad T=\frac{\hbar}{6}t\to+\infty.
\ee
The complex conjugated periods of the Weierstarass function    and the phase 
$\theta(T)=$  slowly depend on time and in the leading approximation do not depend on $\hbar$  \cite{Boutroux}. 

The periods are determined by the Boutroux condition \cite{Boutroux} (see (\ref{B}) below) which happens to be identical to the integral formulation of the Darcy law \eq{22},\eq{BBB} and is given by   \eq{periods}.  Eq.\eq{u1} valid at complex time in a sector bounded by   Stokes lines  $\mbox{Arg}\,t=\pm \frac{\pi}{5}$.


A value of the phase depends on the solution chosen. If one choses to characterize the \Pe-function by the constant $C$ in \eq{exp1}, one can also characterize the phase by $C$.  Relation between $C$ and $\theta$ links   asymptotes of \Pe-function at $T<0$ and $T>0$. It is called {\it connection formula} and obtained  in  Refrs. \cite{Kapaev-Kitaev, Kapaev2001}. Connection formula could be best expressed through the monodromy data 
$\Phi$ - a (locally) time independent constant  defined below in Sec.\ref{MKB}\be\la{phase}
\pi\theta=\Phi\bar\omega+\bar\Phi\omega,
\quad
C=-\IM e^{-\ii\Phi},\quad 2\RE e^{\ii\Phi}=-1.
\ee

The phase $\theta$ is understood through initial data for \Pe equation.  As it follows from (\ref{u1}), at  times 
 \be\la{tt}\frac 45 t_n=\theta+(2n+1)(\omega+\bar\omega),\ee
  where $n$ is a large positive  integer,   the \Pe-function reaches the minima $u_t(t_n)=0$ as in Fig.~\ref{Gamba}.   At those points the value of the solution $u(t_n)\to e(T)$,  where $e(T)$ is a real branch point of the Krichever-Boutrox elliptic curve \eq{e1}.

Scale invariance \eq{scale} implies that the periods behave as $|T|^{-1/4}$ \eq{periods}. Taking this into account the formula \eq{u1}
reads
\be\la{u2}
u(t)\approx \left(\frac{\hbar t}{6}\right)^{1/2} \wp \left[\frac 4 5t \left(\frac{\hbar t}{6}\right)^{1/4}-\theta(1)\Big |\,\omega(1),\bar\omega(1)\right],\quad t\to+\infty,
\ee
where $\theta(1),\;\omega(1)$ are just numerical constants.

We will derive all the formulas of the oscillatory regime below. Connection formula (\ref{phase}) requires a more subtle analysis and we do not discuss here.

\subsection{Adiabatic invariant and Boutroux condition}

\Pe equation can be interpreted as mechanics of a particle with a coordinate $u$ moving  in a potential  $-2u^3$, under slowly changing force $\hbar t$. At $t\to -\infty$ the particle with a vanishing  velocity  is located in the unstable position where the force equilibrates the potential as it follows from (\ref{A1}).  From there as time progresses the force no longer equilibrates the particle. The particle moves to infinity, but returns from there at a finite time,  performing an oscillatory motion.

In oscillatory regime the r.h.s. of the \Pe- equation (\ref{P1})  is a  ``slow time"  $T$,  barely changing over many  periods  of oscillations.
Treating the r.h.s. of (\ref{P1})  as a parameter, the equation becomes an identity for the Weierstrass elliptic function $\wp''-6\wp^2=-\frac{g_2}{2}$, where $g_2=-12 T$. This gives  (\ref{u2}),  where the Weierstrass constant  $g_3$ and the phase $\theta$ remain undetermined.  

Adiabatic nature of modulation of oscillations helps to  determine the Weierstrass constant  $g_3$ and the periods of oscillations.

The energy of the oscillatory motion is
\be\la{H}
H=P^2-4u ^3+g_2 u,
\ee
where  $P\equiv u_t=\sqrt{4u^3-g_2 u+H}$ is the momentum of oscillations.

The Weierstarass parameter $g_3$ is a value of the energy (\ref{H}) evaluated on (\ref{tt}):  $H\to -g_3$. 
The adiabatic invariance of the motion $\oint Pdu$ does not depends on time.     At earlier time when the  oscillations do not yet begin - the  adiabatic invariant  is null. It remains null at all time.
\be\la{B}
\oint  \sqrt{4u^3-g_2 u-g_3}du=0.
\ee
The integration in this formula goes over the period of oscillations.

The fact that the adiabatic invariant vanishes  also  means that at early time  the mean value of the energy \eq{H}  
$H_{ |_{ T\to-\infty}}=-g_3\to \left(g_2/6\right)^3=e^3=-8T^{3/2}$ is such that  the curve $4u^3-g_2 u-g_3=4(u-e)(u+\frac {e}{2})^2$  degenerates as the period of oscillations tends to infinity.  

Eq.\eq{B} is called  Boutroux  condition \cite{Boutroux}. It determines the mean energy  $\overline  H=-g_3$ oscillations and the Weierstarass parameter $g_3$ and subsequently determines the periods.  Boutroux condition \eq{B} is equivalent to the condition \eq{BBB} obtained from the weak solution to the Hele-Shaw problem.  
 The calculations and the results  \eq{21},\eq{periods} were described in  Sec.~\ref{G}. 

 Direct integration of the \Pe equation also yields  the Boutroux condition.
First we write the 
\Pe- equation  in the form which emphasizes a  non-conservative  nature of motion
\be\la{D}
H_t=-2\hbar u,\ee
where $H$ is a  time-dependent Hamiltonian of the mechanical system is in \eq{H}. 
Then compute and compare the  means (average over oscillations) of both sides of the equation. In the leading approximation $g_3(T)=g_3(1) T^{3/2}$ depends only on "slow time". Therefore $\p_T \overline {H}\approx -\frac{3g_3}{2T}=12 \frac{3g_3}{2g_2}$.  The integral over the period is $2\cdot 12(\omega+\bar\omega)\frac{3g_3}{2g_2}$. It must be equal to the integral over the period of the 
 r.h.s.  of \eq{D}. With the help of \eq{u2}  we obtain $\overline u=-\frac{1}{\omega+\bar\omega} \int_t^{t+2(\omega+\bar\omega)}u(t)dt=\frac{\eta+\bar\eta}{\omega+\bar\omega}$.  This yields Eq. \eq{phib} which is equivalent to \eq{B}.

In the next section 
we refine  these results through the monodromy of the associated linear problem. In that way the Boutroux condition  emerges exactly in the form of the integrated Darcy law.

\section{Semiclassical analysis of \Pe I transcendent}
 \subsection{Linear equation for the \Pe-wave-function}
A starting point for the semiclassical analysis of the wave-function is a commonly used realization of the
linear problem \cite{Garnier},  equivalent to \eq{XY1} or \eq{L}. Differentiating the first  equation  \eq{L} by time, one expresses $\p_t^3\Psi$ in the second equation through $\Psi$ and $\p_t\Psi$. Then differentiating the result by time again, one expresses $\p_t^2\Psi$ through $\Psi$.  The result can be written as a  linear matrix differential equation  for   the column  vector   $\underline\Psi=\,\left(\Psi, \Psi_t\right)$ treated as a function of space $X$
\bea\la{iso}
i\hbar\p_X\underline\Psi=\mathbb{ Y}\underline\Psi,
\eea
2x2-matrix$\mathbb{Y}$ explicitly depends on $u,\, u_t$ and $t$ (we  express $u_{tt}$ through the Painlev\'e equation)
\bea\nonumber
-i \mathbb{ Y}= u_t\sigma_3+2\left(X-u\right)\sigma_+-2\left((X+2u)(X-u)+3(u^2+T)\right)\sigma_-,
\eea
where  $\sigma_3,\,\sigma_\pm$ are Pauli matrices.

 This linear equation alone, without a reference to  \eq{L},  already determines that $u(t)$ is a  Painlev\'e function, if one demands that    monodromy data does not change in time, while the matrix $\mathbb{Y}$ does. This is the isomonodromic deformation interpretation of the Painlev\'e equation \cite{Jimbo1,Fokas2}.

 The linear equation \eq{iso} has one irregular singular point at infinity, and three turning points where $\mbox{det}\,\mathbb{Y}=0$.  Solutions are entire functions.

 \subsection{Modulated spectral curve}
The eigenvalues  $\pm Y_\hbar$ of the traceless matrix $\mathbb{Y}$  are two branches of a spectral elliptic curve 
\be\la{R}
-Y_\hbar^2=-\mbox{det}\mathbb{Y}=4X^3-g_2X+H.
\ee
 Here $H$ is given by \eq{H}, and, again, $g_2=-12T=-2\hbar t$.
Another form of the curve is  also useful
$$
-Y_\hbar^2=4(X+2\sqrt{-T})\left(X-\sqrt{-T}\right)^2-4\left(u+2\sqrt{-T}\right)\left(u-\sqrt{-T}\right)^2+ u_t^2
$$
It is organized such that at $T\to -\infty$ one easily recovers a degenerate curve of the Hele-Shaw flow \eq{133}. The last two terms vanish in that  limit.

We emphasize that $Y_\hbar$ features fast oscillations with oscillations of  energy $H$. The average over the period of the oscillations $H\to \overline{H}=-g_3$ revials a slowly evolving modulated spectral  curve $Y(X)$.
\subsection{Semiclassical wave-functions}
The spectral curve    determines the  leading order of the semiclassical approximation of the
 wave-function, while  eigenvectors of the matrix $\mathbb{Y}$ determine the sub-leading term \footnote{The leading term of the  WKB  solutions is $\exp{\int^X\left(\frac{1}{\hbar}\dd\Omega_\hbar-(U^{-1}\dd U)_{11}\right)} $, where $U$ is a matrix diagonalizing $\mathbb{Y}:\; U^{-1}\mathbb{Y} U=Y_\hbar\sigma_3$. Each solution corresponds to a branch of $Y_\hbar$.} \cite{Wasow}.Two independent  basis  WKB-functions  correspond to each branch of $Y$
  \be\la{WKB0}
 \psi_\pm^{WKB}(X,t)\sim\pm\left(\frac{X-u}{Y}\right)^{1/2}
 e^{
  - \frac{u_t}{2}\int_X^\infty \frac{dX'}{(X'-u)(\ii Y(X'))}
  }e^{\pm \frac{1}{\hbar}\Omega_\hbar},
  \ee
where $\dd \Omega_\hbar=-\ii Y_\hbar \dd X$.  The formula   holds  away of a $\sqrt \hbar$ vicinity of three turning points where $Y_\hbar$ vanishes. The function $\psi_+$  matches the asymptote of the wave-function  \eq{V} at $X\to\infty$, which is computed  directly  from \eq{iso}
   \be\la{A2}
\Psi(X,t)=\psi_+^{WKB}=  \frac{1}{\sqrt[4]{X}}\left[1-\frac{H/8\hbar}{ \sqrt{X}}
 +\dots\right ] e^{\frac {1}{\hbar} \left[ \frac{4}{5}X^{\frac 52}-t\,X^{\frac 12}\right]}.
 \ee
 In further  approximation we replace the oscillating $H$
by its mean $\overline H$ and call it $-g_3$. 
 We already know that $g_3$ is
determined either by \eq{phib}, or \eq{B}, but we do not assume it at this point.

Semiclassical form of the \Pe-function can be obtained from \eq{WKB0} under requirement that
the WKB basis is locally single valued functions on the double-covering of the curve  (the rhombus in Fig.~\ref{rhombus})  with branch points at infinity and at turning points $e_{1,2}$.

We set $X=\wp(\lambda),\;-\ii Y=\wp'(\lambda)$ uniformmizing the curve  \eq{R} and choose $\lambda$ to be in  the shaded domain  of the rhombus in Fig \ref{rhombus}. Let us also set $u(t)$  
 to be the Weierstrass function $u(t)=\wp(\tau )$  with the same periods as $X$ and $Y$,  where $\tau$ is some function of time. Then, with a help of  the identity $-\frac{\wp'(\tau)}{\wp(\lambda)-\wp(\tau)}=-\zeta(\lambda-\tau)+\zeta(\lambda+\tau)-2\zeta(\tau)$ we compute $- \frac{u_t}{2}\int_X^\infty \frac{dX'}{(X'-u)(\ii Y(X')}= \frac{u_t}{2}\int_\lambda^0\frac{d\lambda'}{\wp(\lambda')-\wp(\tau)}= \frac{u_t}{2\wp'(\tau)}\left(\log\frac{\sigma(\lambda+\tau)}{\sigma(\lambda-\tau)}-2\lambda\zeta(\tau)\right)$.   Another identity reads $X-u=\wp(\lambda)-\wp(\tau)=\frac{\sigma(\tau-\lambda)\sigma(\lambda+\tau)}{\sigma^2(\lambda)\sigma^2(\tau)}$. Combining the pieces  we observe 
 that the WKB -wave-functions are meromorphic only if   $u_t=\wp'(\tau)$.  
 The latter  determines the argument of $\wp$ function as in \eq{u2}
$$
 \tau=\frac 45 t-\theta,
$$
 leaving the phase $\theta$ as a parameter of solution.

Summing up, the WKB solution reads 
 \be\la{WKB}
 \psi_+^{WKB}(\lambda,t)=\frac{\sigma(\tau+\lambda)}{\sigma(\lambda)\sigma(\tau)}e^{-\zeta(\tau)\lambda} \left(\frac{e^{\frac{1}{\hbar}\Omega_\hbar}}{\sqrt{Y}}\right),\quad\lambda\in B, 
 \ee
where $\lambda$ is chosen to be in the shaded  domain $B$ of the rhombus Fig. \ref{rhombus}.  Here $\sigma$ and $\zeta$ are  elliptic functions of Weiertstra{\ss}. We grouped terms such as to emphasize the slow time dependence of the spectral curve $Y(X)$ and fast oscillations due to  ``fast time" dependence of eigenvectors of $\mathbb{Y}$. 

The second component of the WKB basis can be obtained as a result of rotation of \eq{WKB} around infinity  \cite{Grinevich-Novikov} 
\bea
&&\psi_\pm ^{WKB}(\lambda,t)={\rm i} \psi_\mp^{WKB}(e^{{\rm i}\pi}\lambda,t),\quad \lambda\in B \la{SS},\\
&&\psi_+^{WKB}(X e^{2\pi \ii})=e^{\frac{\ii \pi}{ 2}}\psi_-^{WKB}(X)
\eea
At moments $t_n$ in \eq{tt}, the WKB wave-functions are especially simple
\be\la{Wt}
\psi_\pm^{WKB}(X, t_n)=\pm\sqrt[4]{\frac{X-e}{(X-e_1)(X-e_2)}}e^{\pm\frac{1}{\hbar}\Omega_\hbar }.\ee
 
  \subsection{Monodromy data }\label{MKB}
The basis of WKB solution is quasi-periodic
\begin{equation}
\la{BF}
\begin{array}{c}
\Psi_\pm^{WKB}(\lambda+2\omega)=-\Psi_\pm^{WKB}(\lambda)e^{\mp \ii\Phi},
\\\Psi_\pm^{WKB}(\lambda+2\bar\omega)=-\Psi_\pm^{WKB}(\lambda)e^{\pm \ii\bar\Phi}.
\end{array}
\end{equation}
Complex Bloch factor $\Phi$ constitutes the monodromy data.

The fundamental fact about the monodromy data  is that they do not depend on time if $u(t)$ is a  \Pe-function.  The reverse is also true - if $\Phi$ does not depend on time, then $u(t)$ is a \Pe-function.   This statement in full generality was proven in Ref. \cite{Jimbo1}.  For the case of \Pe I equation the time independence of the monodromy data can be seen from the following simple argument:  Wronskians of  \eq{XY1} computed out of pairs of solutions which WKB asymptotes in a given sector  are $\psi_+^{WKB}(\lambda),\psi_-^{WKB}(\lambda)$ and $\psi_+^{WKB}(\lambda),\psi_-^{WKB}(\lambda+2\omega)$ are related by the Bloch multiplier. Both do not depend on 
 time.  In a  scaling solution Bloch multipliers also do not  depend on $\hbar$ - they are two complex conjugated numbers which could be chosen to characterize the \Pe-function. As it has been shown in \cite{Kapaev1989} physical solution is  selected  by the condition $2\RE e^{\ii\Phi}=-1$.

A jump of  WKB wave-functions (\ref{BF})   on branch cuts can be expressed through Bloch factors.  With a help of \eq{SS} we write
$\psi_+^{WKB}(\lambda+2\bar\omega)=\ii \psi_-^{WKB}(-\lambda-2\bar\omega)=-e^{-\ii\Phi}$. The argument $-\lambda-2\bar\omega$ is a rotation of  a point $\lambda$ in the rhombus by $\pi$ around $-\bar\omega$. In the $X$-space it represents a rotation by $2\pi$ around the turning point $e_2$. Thus
\begin{equation}
\la{JUMP}
\begin{array}{c}
\psi_\pm^{WKB}(e_1+Xe^{2\pi\ii})=\ii e^{\pm \ii\bar\Phi} \psi_\mp^{WKB}(e_1+X)\\ \nonumber
\psi_\pm^{WKB}(e_2+Xe^{-2\pi\ii})=\ii e ^{\mp \Phi} \psi_\mp^{WKB}(e_2+X),\\ \nonumber
\end{array}
\end{equation}
Together with \eq{SS} these equation describe jumps of the WKB-basis on branch cuts. The factors $ s_1=-\ii e^{-\ii\bar\Phi}$ and $s_{-1}=-\ii e^{\ii\Phi}$ are often called  Stokes multipliers \cite{Kapaev1989}.

\subsection{Monodromy data Krichever-Boutroux condition}
 An explicit form of the wave-functions (\ref{WKB}) contains all essential information about \Pe-function. We have  
\bea\la{M1}
-\ii\bar \Phi&&= 
2\left(\omega\zeta(\tau)-\eta\tau\right)-\frac 1\hbar\left(\Omega_\hbar(\lambda+2\omega)-\Omega_\hbar(\lambda)\right)\\
&&=2\omega\left(\zeta(\tau)-\frac 35 \frac{H}{\hbar}\right)+2\theta\eta \nonumber,
\eea
where we used 
\be\la{B1}
\Omega_\hbar(\lambda+2\omega)-\Omega_\hbar(\lambda)=\frac 45  g_2\left(\eta+\frac {3 H}{2g_2} \omega\right)
\ee
and standard formulas $\sigma(\lambda+2\omega)=-e^{2\eta(\lambda+\omega)}\sigma(\lambda), \;\zeta(\lambda+2\omega)=\zeta(\lambda)+2\eta$.
 
 Taking a real part of \eq{M1} we obtain $\RE\oint \dd\Omega_\hbar= \RE(\Omega_\hbar\left(\lambda+2\omega+2\bar\omega)-\Omega_\hbar(\lambda)\right)=
2\hbar(\omega+\bar\omega)\left(\zeta(\tau)-\frac{\eta+\bar\eta}{\omega+\bar\omega}\tau\right) -\hbar\,\IM \Phi$. Averaging over one time period cancels the oscillatory term in the r.h.s.. Hence 
\be\la{barO}\RE\oint \dd\overline{\Omega}_\hbar= -
\hbar\,\IM \Phi.
\ee
The limit $\hbar\to 0$ yields   the Krichever-Boutroux curves  \eq{22}.

Another form of this result is a
 correction to the mean energy: 
 \be\la{barH}\overline H=-g_3+\frac{5\hbar}{3}\frac{\IM\Phi}{\omega+\bar\omega}.
 \ee
 It follows from averaging  the real part of \eq{M1} over the time-period.

Finally, the relation between the phase and the monodromy data \eq{phase} $\pi\theta=\Phi\bar\omega+\bar\Phi\omega$ follows from \eq{M1} - multiply \eq{M1} by $\bar\omega$, take the imaginary part and use  the Legendre relation $\eta\bar\omega-\bar\eta\omega=\frac{\pi \ii}{2}$. 

  One checks directly that in the leading order \eq{M1} is consistent with \eq{D} only if  the monodromy data are time independent. Differentiate  \eq{M1} by the  fast time and  take  into account that all terms in \eq{M1} which  do not scale with a slow time.  Eq.\eq{D} averaged  follows.

\section{Semiclassical  wave-function}
\begin{figure}[h!]
\begin{center}
 \includegraphics[width=2.5in]{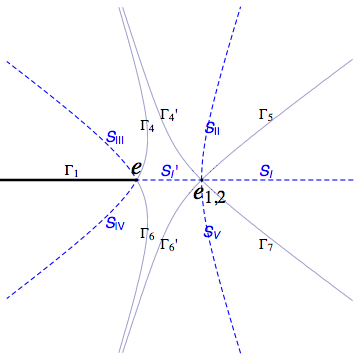} 
\includegraphics[width=2.5in]{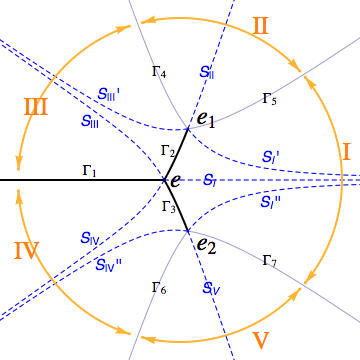}
\caption{  Stokes lines $\IM\Omega=0$ (blue-dashed lines)   before (left panel)  and after (right panel) transition. Stokes -level  lines  are black solid lines.  Before the transition four Stokes  lines emanate from the double point. Stokes phenomena occurs only on $\Gamma_1$ representing the fluid boundary.   After the transition the double point splits into two branch points, each emanating three Stokes lines.   Stokes phenomena occur on  shocks $\Gamma_3, \Gamma_2$ where the \Pe-wave-function accumulates zeros.}
\label{physicalplane1}
\end{center}
\end{figure}

Now we are ready to discuss  the WKB-approximation of the  physical \Pe-I wave-function determined by the condition \eq{V}. The WKB form depends on the sector $S$ bounded by two consecutive  Stokes lines $\IM\Omega(X)=0$ (blue-dashed lines on Fig.~\ref{physicalplane1}) and a real negative axis.  
\be\la{PH1}
\Psi(X)=A_{+}(S)\,\psi_+^{WKB}(X)+A_{-}(S)\,\psi_-^{WKB}(X),\, X\in S.
\ee
 
 Away from Stokes level-lines where $\RE\Omega(X)=0$,  only one term in \eq{PH1} is relevant. It is the dominant term: either $\psi_+^{WKB}$ if $\RE\Omega>0$, or $\psi_-^{WKB}$ if $\RE\Omega<0$, provided that its coefficient does not vanish. If the coefficient of the dominant term vanishes, the WKB approximation is given by the recessive term $\psi_+^{WKB}$ if $\RE\Omega<0$, or $\psi_-^{WKB}$ if $\RE\Omega>0$.  On Stokes-level  lines, both terms in \eq{PH1} oscillate and are comparable.
 
  If both coefficients in \eq{PH1} are non-zero  in some  sector, the physical wave-function has a set of zeros located on a Stokes-level line bisected  this sector. Zeros   accumulate toward  turning points -  end points of the lines. 
  In Hele-Shaw hydrodynamics  (Sec. \ref{S}), some of these particular Stokes-level lines are located in the fluid. They   appear as shocks.

  In order to complete the semiclassical analysis we have to find  the Stokes coefficients $A_\pm(S)$ for the physical \Pe-functions. They are expressed through Bloch multipliers $\Phi$ and   depend neither on time nor on $\hbar$.  

%
%
%
%
%

%


\subsection{Stokes coefficients}
In further analysis, Stokes level-lines (also called anti-Stokes lines), Stokes lines  play an important role.  They have  been already discussed in Sec.\ref{SS}. We recall the major facts.

Stokes level-lines  are curved  lines determined by the condition $\RE\Omega(X)=0$. On these lines the basis  WKB-functions oscillate. There are seven level lines as indicated on the right panel in  Fig. \ref{physicalplane1}.  Three of them  ($\Gamma_{1,2,3}$) are distinct:  $\RE\Omega$ approaching to these lines stays positive on both sides. On other lines $\RE\Omega$, changes sign. 
Being positive in the sector bounded by $\Gamma_{7,3,2,5}$),  $\RE\Omega$ becomes negative in adjacent sectors bounded by $\Gamma_{5,4}$ and in the sector bounded by $\Gamma_{6,7}$,  and positive again in the sectors  bounded by $\Gamma_{4,2,1}$) and   $\Gamma_{1,3,6}$).

Stokes lines are the lines where $\IM\Omega(X)=\mbox{const.}$ and connected to a branch point. There are nine of them as shown in the right panel of   Fig. \ref{physicalplane1}. 
All Stokes lines end at infinity. 

%
%

 The semiclassical  wave-function is analytic between adjacent Stokes  lines separated by a Stokes level line, but may be discontinuous passing through a Stokes line.
 
Condition  \eq{V} requires  that at large $X$ the  semiclassical approximation suffers a discontinuity  only at  a fluid boundary $\mbox{Arg}\,X=\pi$ where the wave-function  rises as $|X|\to\infty$.  This means that  on the both sides of the fluid boundary the wave-function possesses a component  $\psi_+^{WKB}$  which happens to be dominant in the sectors bounded by $\Gamma_1,\Gamma4$ and $\Gamma_1,\Gamma_6$.  In other words $A_+\neq 0$ in these sectors. We set  it to be 1.       Passing the Stokes-level line $\Gamma_4$ the component $\psi_+$ becomes recessive. Therefore $A_-=0$ to the right of $\Gamma_4$, and thus holds everywhere  in the sector bounded by $S_{\rm I}'$ and $S_{\rm III'}$.  The symmetry with respect to complex conjugation insures that $A_-=0$ also in the sector bounded by 
$S_{IV'}$ and $S_{I''}$. Thus 
 \be\la{111}
\Psi\approx\psi_+^{WKB}\, \,\mbox{between    $S_{\rm IV'}- S_{\rm I''}$ and $S_{\rm I'} - \,S_{\rm III'}$}.
\ee

In the sectors adjacent to $\Gamma_1,\Gamma_2,\Gamma_3$  the WKB approximation of $\psi_+^{WKB}$  is dominant. There is no information so far about $A_-$ there. 
Monodromy property of WKB functions determines this coefficient.

Passing  through the Stokes-level $\Gamma_1$, the  term $A_- \psi_-^{WKB}$ becomes dominant. With a help of \eq{SS} we evaluate it at a remote point $X_1=|X_1|e^{\ii( \pi+\delta)},\;|X_1|\to\infty$ on the lower side of the  $\Gamma_1$ ($\delta>0$). We have  $-\ii A_-\psi_+ (\ii |X_1|e^{\ii(-\pi+\delta)})$. But the coefficient of the dominant term is always 1. Thus 
 $A_-=i$  in a sector between $\Gamma_1$ and $S_{\rm III}$.

 Similar arguments determine  the Stokes multipliers  between Stokes lines $S_{\rm III},\,S_{\rm III'}$ and $S_{\rm I},\,S_{\rm I'}$. Summing up
 \be\la{114}
\Psi=
\left\{
\begin{array}{lr}
\psi_+^{WKB}+i \psi_-^{WKB},\,  \mbox{between $\Gamma_1,\,S_{\rm III}$},\\
\psi_+^{WKB}-\ii e^{\ii\bar\Phi}\psi_-^{WKB},\,  \mbox{between $S_{\rm III'},\,S_{\rm III}$},\\
\psi_+^{WKB}+\ii e^{\ii\bar\Phi}\psi_-^{WKB},\,  \mbox{between $S_{\rm I'},\,S_{\rm I}$},\\
\psi_+^{WKB},\,  \mbox{between $S_{\rm I'},\,S_{\rm III'}$}.
\end{array}
\right.
\ee
The WKB-approximation of the physical wave function in sectors in the lower half plane is achieved by complex conjugation.
 A special feature of this solution is that  at infinity, everywhere in a fluid (a complex plane  cut  by $\Gamma_1$) $\Psi\approx\psi^{WKB}_+$ and  Eq. \eq{V} holds.
 
 Situation is different at $T<0$. Configuration of Stokes lines and Stokes-level lines is depicted in the left panel of Fig. \ref{physicalplane1}. In this case the only Stokes-level line where $\Re\Omega$ stays positive  on both sides is $\Gamma_1$. Similar analysis yields a very different result
\be\nonumber
\Psi=
\left\{
\begin{array}{lr}
\psi_+^{WKB}+i \psi_-^{WKB},\,  \mbox{between $\Gamma_1,\,S_{\rm III'}$},\\
\psi_+^{WKB},\,  \mbox{elsewhere\, in\, the\, upper\, half\, plane}.
\end{array}
\right.
\ee


\section{Semiclassical wave-function and the Hele-Shaw flow}

\subsection{ Darcy Law as an average over the period}
The Darcy law  follows from the semiclassical wave-function.   As it has been described in Sec. \ref{A0} one recovers the hydrodynamics by averaging oscillating quantities over the period.  Let us see how does the wave-function change over the time period $2(\omega+\bar\omega)$.   

Away from sectors adjacent to $\Gamma_2,\,\Gamma_3$ and a fluid boundary $\Gamma_1$ the WKB wave-function is equal to  $\psi_+^{WKB}$.  Let us compute an average $\hbar\log\Psi$ over  fast oscillations. 

First we notice that the term placed  in \eq{WKB} outside of the brackets $\frac{\sigma(\tau+\lambda)}{\sigma(\lambda)\sigma(\tau)}e^{-\zeta(\tau)\lambda}$ is time-periodic. The only change over the period comes from the slowly varying part $Y^{- 1/2}e^{\frac{1}{\hbar}\Omega}$  in  \eq{WKB}.  Thus the leading contribution into $\hbar \log\Psi$ is $\Omega$. Eqs. \eq{Y} recover the weak solution of the Hele-Shaw problem.
%

\subsection{Locus of zeroes of the physical wave function and shocks}
Physical wave-function oscillates on Stokes level lines. Three Stokes level lines $\Gamma_{1,2,3}$ are  special. There, the oscillating WKB wave function possesses two oscillatory terms  \eq{114}, and, therefore, has zeroes. 
 Position of zeroes $X_k$ on these level lines is determined by  Bohr-Sommerfeld conditions
\be\la{BS}
\Omega_\hbar(X_k)=
\left\{
\begin{array}{lr}
\ii \hbar \pi(k - \frac 14), \quad X_k\in\Gamma_{1},\\
\ii \hbar (k\pi+\frac \pi4+\frac 12\overline\Phi), \quad X_k\in\Gamma_{2},\\
\ii \hbar (k\pi+\frac \pi4-\frac 12\Phi), \quad X_k\in\Gamma_{3},
\end{array}
\right.
\ee
where we evaluate $\Omega_\hbar$ from the above of the cuts.
We observe a simple pattern selecting the locus of zeros. These are the lines on which $\RE\Omega$ vanishes being positive at both sides as  in \eq{ad}.

Crossing these lines the continuous wave-function experiences  an abrupt change  on a scale vanishing at $\hbar\to 0$.   In that limit the wave-function suffers a  discontinuity.  The height function, potential, etc., \eq{Y} obtained by differentiation of the wave-function follow suit. In the hydrodynamic  limit $\hbar\to 0$ the lines of accumulating zeros in the fluid  appear as viscous shocks, each zero  appear as micro-vortex as has been discussed in  (Sec.\ref{G}).

\section{Discussion and conclusions}
We have demonstrated that viscous shocks in singular Hele-Shaw flow are intimately related to non-linear Stokes phenomena of \Pe I - equation. Stokes-level lines in a complex plane  accumulating zeros of the \Pe-wave-functions evolve as viscous shocks.

Hydrodynamics quantities have a transparent interpretation in terms of evolution of semiclassical approximation of the physical \Pe - function. In particular, each zero of the holomorphic  \Pe-wave-function is understood as a vortex in the fluid.

Semiclassical approximation of \Pe-wave-function at $T>0$ features fast time oscillations and also  fast spatial oscillations along shocks. Hydrodynamics can be read from these objects either through averaging them over periods of oscillations \eq{Y}, or, alternatively, by considering discrete moments of time  \eq{tt}  corresponding to half-period where $u_t=0$. This sequence of times depends on monodromy data, while hydrodynamics does not
$$\begin{array}{lll}
T_n& = & \left( \frac {5\hbar}{12}\right)^{4/5}\Big((n+\RE\Phi)\RE\omega(1)-\IM\Phi\,\IM \omega(1)\Big)^{4/5} \\
& \approx & \left( \frac {5\hbar}{12}\cdot 0.7427\right)^{4/5}\left(n+\RE\Phi-\IM\Phi\cdot 0.8173
\right)^{4/5}.
\end{array}
$$
Here $n$ is large positive integer.
At these moments the wave-function reads
\be\nonumber
|\Psi^{WKB}(X)|^2\sim \left[\frac{X-e(T)}{X^2 + e(T) X + \frac{g_3}{4 e(T)}} \right]^{\frac 14}e^{\frac{4}{5\hbar}\IM (XY-2T \phi(X))}
\ee
where
$$Y(X)=\sqrt{-4X^3+12 T X+g_3},
$$
$$
 \phi(X)=6\int_{e(T)}^X\left( X'+\frac {g_3}{8T} \right)\frac{dX'}{Y(X')}.$$
The numerical value of $g_3$ and $e(T)$ are given in (\ref{21}) and (\ref{e1}) respectively by
\begin{equation*}
g_3\approx -7.321762\, T^{3/2},\quad e(T)\approx -0.553594\sqrt{T}.
\end{equation*}
Eqs. \eq{Y} yields hydrodynamics results for potential and velocity of the flow supplemented by semiclassical corrections.

Discrete nature of time and discrete structure of shocks supports a ``corpuscular" interpretation of singular flow. One may associate a space between zeros in a  shock line with a particle aggregating on  a  shock. A particle arrives to a shock  every   time interval equal to   $\hbar^{4/5}$.

Finally we comment that the present analysis does not discriminates within  a one-parametric  family of physical solutions characterized by the Bloch factor $\Phi$. We conjecture that a unique solution will be selected by the requirement  that \Pe-function is   associated with  some stochastic process.  This requirement may select $\Phi$. An interesting candidate is $\IM\Phi=0$. Then the Bloch factor is a pure phase. According to \eq{phase} $\Phi=\pi/6$ and $|C|=\sqrt{3}/{2}$.
\section*{Acknowledgements}
S.-Y L. was supported by Sherman Fairchild senior researcher fellowship. P. W. was supported by NSF DMR-0540811/FAS 5-2783, NSF DMR-0906427, MRSEC under DMR-0820054 and the FASI of the Russian Federation under contract 02.740.11.5029.  P. W. and R. T. also acknowledge the USF College of Engineering Interdisciplinary Scholarship Program and the support of USF College of Arts and Sciences.  We thank A. Its and A. Kapaev for helpful discussions and are  especially grateful to I. Krichever.

\section*{References}


\def\cprime{$'$} \def\cprime{$'$} \def\cprime{$'$} \def\cprime{$'$}
  \def\cprime{$'$}

\end{document}